\documentclass[prl,twocolumn,floatfix,showpacs,superscriptaddress]{revtex4}
\usepackage{graphics}
\usepackage{epsfig}
\usepackage{times}

\begin{document}
%\draft
\preprint{HEP/123-qed}
\title{Dimensional tuning of electronic states under strong and frustrated interactions}
\author{Chisa Hotta}
\affiliation{Kyoto Sangyo University, Department of Physics, Faculty of Science, Kyoto 603-8555, Japan}
\author{Frank Pollmann}
\affiliation{Max-Planck-Institut f\"ur Physik komplexer Systeme, 01187 Dresden, Germany}
 %\footnote{\vspace*{-10mm} electric address: chisa@phys.aoyama.ac.jp} 

\date{\today}
%%{submitted in }
\begin{abstract}
We study a model of strongly interacting spinless fermions on an anisotropic triangular lattice. At half-filling and the limit of strong repulsive nearest-neighbor interactions, the fermions align in stripes and form an insulating state. When a particle is doped, it either follows a one-dimensional free motion along the stripes or fractionalizes perpendicular to the stripes. The two propagations yield a dimensional tuning of the electronic state. We study the stability of this phase and derive an effective model to describe the low-energy excitations. Spectral functions are presented which can be used to experimentally detect signatures of the charge excitations. 
\end{abstract}
\maketitle
%\begin{multicols}{2}
\narrowtext 
Electronic properties of metals are essentially understood on the basis of band theory 
which straightforwardly reflects the actual crystal geometry. In certain strongly correlated materials, however, the electronic interactions can significantly modify the spacial geometry of the wave functions (metallicity). For example, in high-$T_c$ cuprates, the charges on the square lattice may align in stripes and the metallic properties become one dimensional (1D) \cite{emry}. Also in manganites, the strong correlations of spin and orbital degrees of freedom lead to orbital orders in which the metallicity is confined to a particular spatial direction \cite{kugelkhom}. Here the density of states reflects correlation effects and causes an effective modulation of the lattice geometry and thus leads to a change of the intrinsic thermodynamical and transport properties. A recently proposed way to tune the effective geometry of a lattice is to introduce frustrations in the electronic interactions. This is seen in a model of spinless fermions (i.e., fully spin polarized electrons) on a triangular lattice with strong nearest neighbor repulsion which is referred as a "pinball liquid". Part of the fermions form a Wigner crystal on one of the three sublattices. The other fermions form a metallic state on the remaining sites along a honeycomb lattice \cite{chisa}. 
%%Similar behavior is found in the hard core bosonic model as a supersolidity\cite{supersolid}. 
\par
In this Letter, we investigate a tuning of the effective dimensionality of the metallicity due to {\it partially frustrated and strong interactions}. We focus on an anisotropic triangular lattice system and start from an insulator at half-filling with striped charge order. Doping of particles yields a 1D metallic state as in the case of the square lattices in cuprates. In addition, we find another non-trivial propagation in which the doped charges fractionalize and change the stripe geometry. 
%
%%%%%%%%%%%%%%%%%%%%%%%%%%%%%%%%%%%%%%%%%%%%%%%%%%%%%%%%%%%%%%%%%%%%%%%%%%%%%%
%*%*%*%*%*%*%*%*%*%*%*%*%*%*%*%*%*%*%*%*%*%*%*%*%*%*%*%*%*%*%*%*
%*%*%*%*%*%*%*%*%*%*%*%*%*%*%*%*%*%*%*%*%*%*%*%*%*%*%*%*%*%*%*%*
%*%*%*%*%*%*%*%*%*%*%*%*%*%*%*%*%*%*%*%*%*%*%*%*%*%*%*%*%*%*%*%*
%
\par
\begin{figure}[tbp]
\begin{center}
\includegraphics[width=7cm]{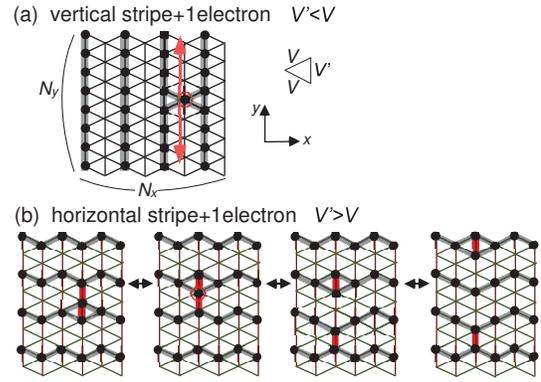}
\end{center}
\caption{(Color online) Representative ground-state configurations of the Hamiltonian in Eq.(1) on the anisotropic triangular lattice in the limit of strong correlations ($t/V=0$) with one particle doped at half-filling. Panel (a) show the case on a vertical stripe order at $V' < V$ where an added particle can move only along the stripes of empty sites. 
Panel (b) shows the horizontal stripe at $V' > V$, where the fractionalization is illustrated by the separation of the two bold vertical bonds in the $y$-direction. }
\label{f1}
\end{figure}

We focus only on the charge degrees of freedom and start from a basic $t$-$V$ model Hamiltonian which reads
\begin{equation}
{\cal H}_{t-V} =\sum_{\langle i,j \rangle} t_{ij}\left(c^\dagger_i c^{\vphantom{\dagger}}_j + {\rm H.c.}\right) +  \sum_{\langle i,j \rangle}V_{ij} n_i n_j. 
\label{tvham}
\end{equation}
Here  $c^{\dagger}_j$ ($c^{\vphantom{\dagger}}_j$) are creation (annihilation) operators of fermions and $n_j$ (=$c^\dagger_j c^{\vphantom{\dagger}}_j$) are number operators. The interactions act only between neighboring sites $\langle ij \rangle$. Anisotropies of the hopping amplitudes and repulsion strengths are given by $(t_{ij},V_{ij})=(t',V')>(0,0)$ for vertical bonds (along the $y$-axis in, e.g., Fig.~\ref{f1}) and $(t,V)>(0,0)$ for the remaining bond directions. We start from the limit of strong anisotropic interactions, i.e.,  $t/V= 0$, $t/V'= 0$, $t/|V'-V|=0$, and assume for simplicity that $t'=t>0$. The ground state at half-filling has an insulating charge order which minimizes the repulsive nearest-neighbor interactions \cite{nakagawa}. 
We consider the effect of doping either a particle or a hole into the insulating state. 
Depending on the anisotropy of the interactions, we find two different scenarios. The case $V'< V$ is depicted schematically in Fig.~\ref{f1}(a). The ground state at half-filling has a vertical stripe-order. An added particle can move freely along a line of empty sites without changing the potential energy of $4V$. All processes of moving orthogonal to the stripes are prevented by an increase of potential energy by $2(V-V')$. This implies that the added particle is confined to a 1D motion along the $y$-axis with a dispersion $-2t\cos(k_y)$ for wave-number $k_y$. Doping of one hole decreases the potential energy by $2V'$ and yields a spectrum identical to that of the added particle. For the case $V' > V$, the ground state at half filling has a high degeneracy. The ground states are formed by isolated configurations which have stripes perpendicular to the anisotropic bonds with a $2^{N_x}$-fold degeneracy \cite{nakagawa}. Figure~\ref{f1}(b) shows the horizontal stripe configuration as one example of degenerate ground state configurations. An added particle can propagate along the stripes as in the former case. However, an additional propagation can take place now. The doped particle interacts with the stripes in the vertical direction by two $V'$-bonds. These two bonds can separate orthogonally to the stripes \emph{without} increasing the potential energy. Each of them is carrying a charge of $-e/2$. These local defects are in the following regarded as fractional charges. As long as we remain in the limit of strong interactions, i.e., assume $t/V = 0$, the fractional charges are deconfined. When a particle is removed, a similar picture arises from the fractionalization of a hole. 
\par
\begin{figure}[tbp]
\begin{center}
\includegraphics[width=8.5cm]{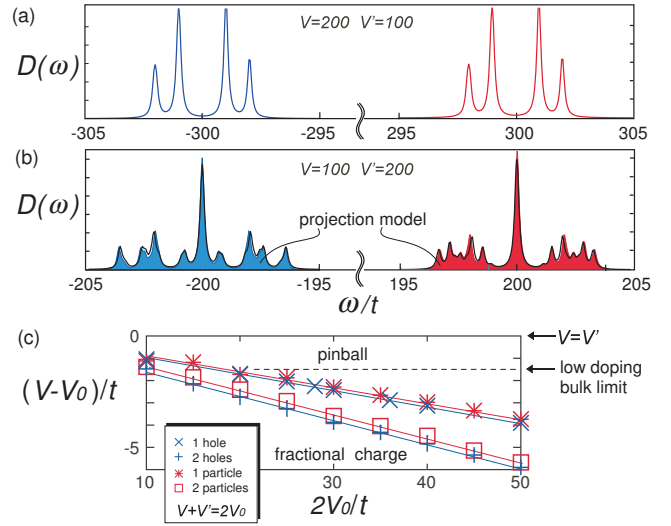}
\end{center}
\caption{(Color online) Numerical results on a $N_x\times N_y=4\times 6=24$-site cluster. 
DOS of Eq.~(1) with $t=t'=1$ for (a) $(V,V')=(200,100)$ and (b) $(V,V')=(100,200)$. 
Panel (b) compares the DOS of Eq.~(1) for $V'=200, V=100$ (bold line) with the effective model (shaded region) of Eq.~(2) on top of each other. (c) shows the ground state phase diagram. The phase boundary is shown for different doping levels. In the bulk system the phase boundary at low doping limit approaches the broken line, where we have fractional-charge phase at $V-V_0 < -\frac{3t}{2}$. 
}
\label{f2}
\end{figure}

In the above discussion, we have neglected any quantum fluctuation in $t/V$. Next, we will investigate how quantum fluctuations affect the ground state and calculate the low-lying excitations. We numerically diagonalize the Hamiltonian in Eq.~(\ref{tvham}) exactly on finite size clusters and derive a strong-coupling effective model by a perturbative approach. We start by calculating the local density of states (DOS), $D^{\pm}(\omega)=\sum_k A^{\pm}(k,\omega)$ of the full Hamiltonian in Eq.~(\ref{tvham}) by means of the Lanczos continued fraction method \cite{lanczos}. Here, $A^{\pm}(\mathbf k, \omega)$ is the spectral function from adding ($+$) and removing ($-$) a particle.  We use a $N=N_x\times N_y=4\times 6$ site cluster with periodic boundary conditions. Figs.~\ref{f2}(a) and \ref{f2}(b) show the DOS for $V'<V$ and $V'> V$ in the strong coupling region with $V+V'\equiv 2V_0=300$. The spectrum of the first case ($V'< V$) in Fig.~\ref{f2}(a) has a width of $4t$ and corresponds exactly to a 1D dispersion in a single-particle picture. For the latter case ($V'> V$), we find instead an incoherent broad structure in Fig.~\ref{f2} (b). The bandwidth is $\approx$$8t$ and thus twice as big as in the former case. This indicates the existence of a collective excitation of fractionalized charges.   
\par
In the following, we focus on the interesting case in which fractionalization of charge occurs. We begin by estimating the parameter range for which the fractional-charge phase sustains. 
In this phase, the ground-state energy depends on the interaction in a specific manner, i.e., we find $E \sim \frac{V}{2}N-(2V+4t)$ and $\frac{V}{2}N+(2V+2V'-4t)$ for one hole and one particle doping, respectively. The first term originates from the stripes and the second term is the interaction energy of the doped charges. On the other hand, a pinball liquid phase is found for $V'\sim V$ (i.e., $V_0 \sim V$) which also shows linear $V$-dependence in energy as discussed in Ref.~\cite{nakagawa}. We distinguish the phases by their different linear slopes in the energy $E(V)$. The phase boundaries are then identified for up to two particle/hole doping ($\sim 8\%$ doping) as shown in Fig.~\ref{f2}(c). From these results we can expect that the fractional-charge phase remains stable for a large range of anisotropic interactions even if the doping level is increased. Note that if we increase the system size towards the bulk limit, the phase boundary for the doped particle number of $O(1)$ will asymptotically approach the solely $t$-dependent line at $V-V_0=-\frac{3t}{2}$. We thus conclude that the physics we discuss is relevant for a large range of interaction strength and different doping levels. 
 
%*%*%*%*%*%*%*%*%*%*%*%*
%*%*%*%* fig 1 *%*%*%*%*
%*%*%*%*%*%*%*%*%*%*%*%*
%*%*%*%*%*%*%*%*%*%*%*%*
%*%*%*%* fig 2 *%*%*%*%*
%*%*%*%*%*%*%*%*%*%*%*%*
%*%*%*%*%*%*%*%*%*%*%*%*
%*%*%*%*%*%*%*%*%*%*%*%*
%*%*%*%*%*%*%*%*%*%*%*%*
%
%*%*%*%*%*%*%*%*%*%*%*%*
%*%*%*%* fig 3 *%*%*%*%*
%*%*%*%*%*%*%*%*%*%*%*%*
\begin{figure}[tbp]
\begin{center}
\includegraphics[width=8cm]{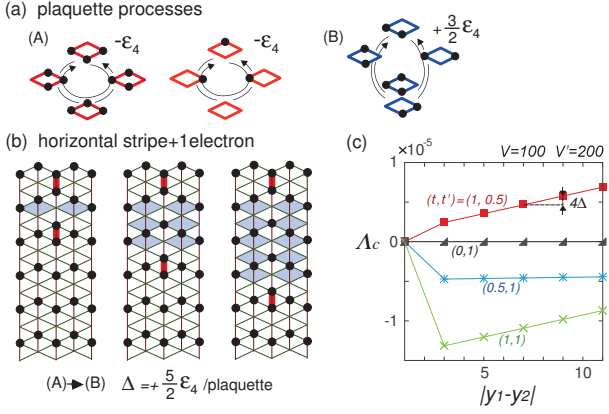}
\end{center}
\caption{(Color online) 
(a) Fourth order ring exchange processes along the plaquettes; (A)/(B) lowers/raises the energy by ($-\epsilon_4$)/($+\frac{3}{2}\epsilon_4$) (both the clockwise and anti-clockwise processes are counted). 
(b) The case when one particle is added. The shaded plaquettes between the two vertical bonds are the plaquettes replaced from (A) to (B) when the bonds are separated by the distance $|y_1-y_2|$. (c) Energy correction $\Lambda_c$ including all the perturbation processes up to fourth order as a function of $|y_1-y_2|$ for several parameter choices. The gradient is $2\Delta=5\epsilon_4$, which comes from the conversion of (A) to (B) plaquettes. 
}
\label{f3}
\end{figure}
%*%*%*%*%*%*%*%*%*%*%*%*
%*%*%*%*%*%*%*%*%*%*%*%*
%*%*%*%*%*%*%*%*%*%*%*%*
\par
Next, in order to understand in detail how the fluctuations affect the low-energy excitation, we perform a perturbative treatment in the limit of strong couplings. We begin with the half-filled case which has a degenerate ground-state manifold of stripe configuration and include the perturbative processes in $t/V$, $t/V'$ and $t/|V-V'|$ up to fourth order. The first contribution which lifts the degeneracy are the fourth-order exchange processes around plaquettes as shown in Fig.~\ref{f3}(a). 
We distinguish between (A) plaquettes with odd number of particles (i.e., either one or three) and (B) plaquettes with two particles. These processes give energy corrections of $-\epsilon_4$ and $+\frac{3}{2}\epsilon_4$ per plaquette, respectively, where $\epsilon_4=2t^4/\big((2V'-V)^2V'\big)>0$. The different signs originate from an exchange of fermions. The conversion of one plaquette from (A) to (B) increases the energy by $\Delta=\frac{5}{2}\epsilon_4$. Then, at half-filling, the afore mentioned degeneracy of stripes at $t/V' =0$ is lifted and the horizontal stripe filled with (A) plaquettes becomes a unique ground state. The lowest excitation energy is $N_y \Delta$ which results from a collective rearrangement of charges to another stripe \cite{nakagawa}. 
In the fractional-charge phase near half-filling, those plaquette-exchange processes lead to a confinement. This can be seen directly in Fig.~\ref{f3}(b); the separation of fractionalized charges converts (A) plaquettes to (B) plaquettes. The confinement potential $\Lambda_c(|y_1-y_2|)$, which sums up all the possible perturbative processes up to fourth order (excluding a constant shift), is shown in Fig.~\ref{f3}(c). Here $y_1$ and $y_2$ are the $y$-components of the locations of the two fractions. The main feature is a linear contribution of $4\Delta$ per two lattice spacings (for $|y_1-y_2|>3$) which arises from part of the fourth-order processes where the conversion of (A) to (B) plaquettes takes place. The above findings allow us to introduce the following effective Hamiltonian
%*%*%*%*  footnote *%*%*%*%*%
\footnote{In the calculation we added a constant energy term in Eq.(2) to adopt the potential level of the non-fractionalized state to be zero. 
The projected basis for the doped system in Eq.(\protect\ref{prham}) includes $2^{N_x}$-degenerate stripe configurations. When the two fractions separate over the distance $N_y$, a mixing occurs between different stripe configurations. This is prevented in the thermodynamic limit by finite confinement potential. In finite systems, however, mixing easily occurs which hinders the clarification of bulk properties. To avoid this artificial mixing effect, we consider only horizontal stripe configurations in most of the calculations except in the shaded spectrum of Fig.~4(d). 
}
%*%*%*%*%*%*%*%*%*%*%*%*%*%*% 
to describe the low energy excitations of the doped system for $V'\gg V \gg t,t'$: 
\begin{equation}
{\cal H}_{\rm pr} = \sum_{\langle i,j \rangle} {\cal P}\left( t_{ij} c^\dagger_i c^{\vphantom{\dagger}}_j + {\rm H.c.}\right){\cal P} + \Delta(N_{\text (B)}-N_{\text (A)}). 
\label{prham}
\end{equation}
The projector ${\cal P}$ projects onto the manifold of allowed configurations which minimize the interaction term of Eq.~(1). The projected configurations are connected by first order in  $t_{ij}$. The second term takes into account the confinement effects which result from the conversions of (A) to (B) plaquettes. The operators $N_{\text (A)}$ and $N_{\text (B)}$ count the number of corresponding plaquettes. 
We validate the effective Hamiltonian by comparing the obtained  electron doped DOS with the one obtained from the Full Hamiltonian Eq.~(\ref{tvham}). As presented in Fig.~\ref{f2}(b), the DOS of Eq.~(\ref{tvham}) for $(V,V')=(100,200)$ in solid line agrees perfectly well with the one from the effective model on the same cluster presented as a shaded region. 
The fact that we have to consider only the manifold of stripe-based configurations reduces the size of the Hilbert space of ${\it {\cal H}_{pr}}$ drastically as compared to the full Hilbert space. Thus we can perform a careful numerical analysis of the effects of the confinement potential. 
\par
\begin{figure}[tbp]
\begin{center}
\includegraphics[width=8cm]{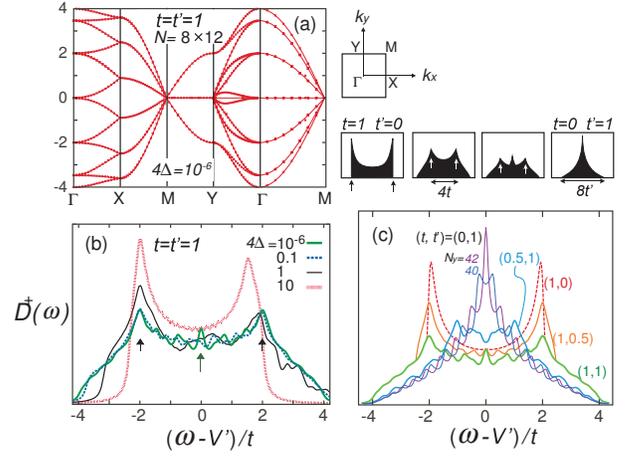}
\end{center}
\caption{(Color online) Panel (a) shows the energy dispersion of the effective Hamiltonian Eq.~(\protect\ref{prham}) for a small confining potential $4\Delta=10^{-6}$ on a $N=8\times 12$ cluster. Panel (b) and (c) show the local density of states $D(\omega)$ for different confining potentials and for different anisotropies in the hopping amplitudes, respectively. We use for the spectrums the $N=4\times 32$ or $8\times 24$ cluster which are large enough to give qualitatively the same results (we checked the finite size effect up to $N_y=42$). Arrows in panel (b) indicate the three intrinsic peak positions. The energy is shifted by a constant $C(\Delta)$ for a clear presentation.   
Inset: Schematic illustration of the variation of the spectrums. Arrows denote the (non-fractionalized) van Hove singularity. 
}
\label{f4}
\end{figure}

Let us first focus on the case, $\Delta \approx 0$. When a particle is added, it moves either along the stripes freely by $t$ or fractionalizes vertically to the stripes by $t'$. This is observed  in the dispersions within the finite cluster. 
Figure~\ref{f4}(a) shows the two-dimensional dispersion for $t\!=\!t'\!=\!1$. The dispersions along $\Gamma-X$ are nearly flat since the $x$-direction motion is sacrificed by the increase of coherence of fractional charges in $y$-direction. There are dispersive degenerate modes with a width of $4t$ along $M$-$Y$. These modes originates from the non-fractionalized portion which can only propagate in $x$-direction. They will be shown later to contribute to the weakened van Hove singularity. The feature of the fractionalization in the $y$-direction is detected along the $X$-$M$ line. We find densely packed branches which are following a cosine dispersion. The bandwidth is $\approx 8t'$ and is almost independent of $N_y$ while the number of modes is $N_y/2$. Thus, the dispersions extrapolate towards a continuum in the bulk limit. The existence of continuum has already been discussed in terms of domain wall excitations in 1D Wigner lattices and is visible in the exact solution \cite{horsch}. The two fractions share the total momentum $k$ with arbitrary ratio so that at each $k$-point the energy can take almost any values of energy within a certain range. 

%*%*%*%*%*%*%*%*%*%*%*%*
%*%*%*%* fig 4 *%*%*%*%*
%*%*%*%*%*%*%*%*%*%*%*%*
%*%*%*%*%*%*%*%*%*%*%*%*
%*%*%*%*%*%*%*%*%*%*%*%*
%*%*%*%*%*%*%*%*%*%*%*%*
%*%*%*%*%*%*%*%*%*%*%*%*
\par
The effect of the confinement potential becomes clear by comparing the DOS obtained for different choices of $\Delta$ (see  Fig.~\ref{f4}(b)). If the confinement is small, we find a three peak structure which results from the superposition of the 1D van hove singularities at $\omega=\pm 2t$ 
and a $\omega=0$-centered broad single peak structure. 
The former singularity is a result of the non-fractionalized propagation in the $x$-direction. 
The latter single peak is due to the fractionalization of charge and is consistent with the previous 1D study \cite{horsch}. The weight of the single peak is suppressed by a very large confining potential $\Delta$. However, even for confinement potentials as large as $\Delta \sim t',t$, we still find broad incoherent structure expanding over the energy range of $|\omega|>2t$. 
This is a clear indication that although the two fractions cannot separate to infinite distance under confinement, the coherence due to $t'$ is still large and they form bound pairs with large spatial extent (usually of order $t'/\Delta$ in units of the lattice constant). 
A similar transformation of the DOS is observed when we consider the case without confinement potential but change instead the anisotropic hopping parameters by hand. Figure~\ref{f4}(c) shows the electron doped spectrum for different ratios of $t'/t$. For $t=0, t'\ne 0$, the fractional charges form a broad single peak structure. When introducing a finite $t$, the two 1D van-Hove singularity peaks appear, and it finally becomes a pure 1D DOS of free fermion along the $x$-direction at $t'=0, t\ne 0$. 
%%*%*%*%*%*%*%*%*%*%*%*%*%*%*%*%*%*%*%*%*%*%*%*%*%*%*%*%*%*%*%*%*
%*%*%*%*%*%*%*%*%*%*%*%*%*%*%*%*%
\par
In summary, we find a state with exotic quasiparticles described by a $t$-$V$ model on the triangular lattice in the presence of strong and anisotropic interactions ($V'>V \gg t,t'$). An added particle or hole decays into bound pairs of fractional charges with large spatial extent. 
Not only the added particle but also the other particles move cooperatively and show collective excitation behavior in the anisotropic direction. 
Even though the system is originally two dimensional, the strong and anisotropic interactions first lead to a 1D like charge ordering at half-filling. The dispersion of the quasiparticle is the combination of the 1D free ($t$) and 1D collective ($t'$) propagations along and perpendicular to stripes, respectively. The effective dimension is tuned from one to two-dimension depending {\it on the anisotropy of $t'/t$, on the perturbative interactions and on the original interaction strength.} 
It is remarkable that such delicate and non-trivial dimensional tuning occurs in such a simple model. 
\par
Fermions and bosons on certain frustrated lattices have shown to behave in a similar manner, e.g., a pinball liquid found previously in the present model \cite{chisa} and a supersolidity in a hard core bosonic model \cite{supersolid} can be compared. 
We find that in the present system, a similar hard-core bosonic model has a diagonal stripe ground state at half-filling instead of the horizontal stripe and it shows similar fractionally charged excitations. These fractionalized excitations are bound by a confinement potential of one-half the magnitude of that of the fermionic model. The different details originate simply from the different statistics, i.e., the absence of a fermionic sign when exchanging two hard-core bosons. 
\par
Finally we note that such dimensional tuning is not limited to the anisotropic triangular lattice. We find similar effects also on the anisotropic kagome lattice at around 2/3-filling.  Such picture in between the classical order and the quantum liquid might more generally be expanded to several other frustrated lattices as well. 
\par
%%*%*%*%*%*%*%*%*%*%*%*%*%*%*%*%*%*%*%*%*%*%*%*%*%*%*%*%*%*%*%*%*
%*%*%*%*%*%*%*%*%*%*%*%*%*%*%*%*%
\par
The two-dimensional charge ordered stripe state originating from strong Coulomb interactions is already an established phenomenon in organic solids \cite{chemrev}. To search for its further interesting possibilities, optics become a powerful tool which aim for the photo-induced phase transitions as in EDO-TTF \cite{koshihara}, $\theta$-ET$_2$X. \cite{iwai} and $\alpha$-ET$_2$I$_3$ \cite{tajima}. 
We expect the highly conducting photo-induced state based on the horizontal stripe in the latter two materials 
to be a good candidate to realize the scenario presented in this letter. 
Fractional charge itself has been discussed in many theoretical articles \cite{horsch,fulde,fpollmann}. 
Although the present quantum fractionalization due to frustration is not exactly the one found in the classical particle picture, it is certainly exotic and is the strong indications for a collective many body effect. Its experimental proof should be awaited. 
\par
We thank Peter Fulde and Karlo Penc for helpful discussions. 
This work is supported by the Max-Planck-Institut f\"ur Physik komplexer Systeme. 
%%*%*%*%*%*%*%*%*%*%*%*%*%*%*%*%*%*%*%*%*%*%*%*%*%*%*%*%*%*%*%*%*
%%*%*%*%*%*%*%*%*%*%*%*%*%*%*%*%*%*%*%*%*%*%*%*%*%*%*%*%*%*%*%*%*
%%*%*%*%*%*%*%*%*%*%*%*%*%*%*%*%*%*%*%*%*%*%*%*%*%*%*%*%*%*%*%*%*
%*%*%*%*%*%*%*%*%*%*%*%*%*%*%*%*%*
%*%*%*%*%*%*%*%*%*%*%*%*%*%*%*%*%*
%%%%%%%%%  REFER  %%%%%%%%%%%%%%%%%%%%%%%%%%%%%%%%%%%%%%%

\end{document}